# Quantification of the Impact of Water on the Wetting Behavior of Hydrophilic Ionic liquid: A Molecular Dynamics Study

*Sanchari Bhattacharjee, Sandip Khan* *

Department of Chemical & Biochemical Engineering

Indian Institute of Technology Patna

Patna, India- 801103.

*Corresponding author: E-mail: skhan@iitp.ac.in

**Abstract**

We have used molecular dynamics simulations to study the effect of water on the wetting behavior and the interfacial structure of ionic liquid (IL) 1-ethyl-3-methylimidazolium boron tetrafluoride[EMIM][BF$_4$] droplets on graphite surfaces which, is a prerequisite for the new IL-based applications in the field of chemical engineering. A slight decrement in the value of the contact angle has been found while adding water up to 20wt%; afterward, the contact angle starts to increase. The non-monotonic behavior of contact angle of IL droplet with the addition of water molecules was examined through several key parameters, such as the interaction energies between the graphite and IL molecules, density distributions, cluster formation, and the number of formed hydrogen bonds (HBs) for the different weight percentages of water molecules. The results indicate that the hydrogen bond network and the cluster formation among the water molecules play a vital role during the transition from IL rich medium to water rich medium of the droplet.

## 1. Introduction

The wetting behavior of ionic liquid at the solid-liquid interface has importance in various applications such as pixel displays fluidic devices, micro-lenses, digital microfluidic devices.[1, 2] There have been multiple studies published recently on the wetting behavior of pure ILs illustrating how the selection of anion, cationic chain length, temperature, type of substrate, application of an external voltage across the solid−liquid interface and surface interaction potential affects the wettability of ILs [3-21]. Understanding the wetting behavior of ILs on different surfaces is purposeful for designing and managing IL related chemical engineering processes.[22-24] Moreover, various technologically motivated research studies attempted to elucidate the role of water molecules in IL/water interfaces, which is yet to understand correctly[11, 25-30].

Pure ILs being extremely viscous in nature, unable to wet the surface efficiently. Thus water content plays an essential role in the wettability of extremely hygroscopic ILs [31] by decreasing its viscosity. Water as an impurity in ILs can have significant effects on their physical properties such as conductivity, solubility, surface tensions, diffusion coefficients, viscosities, and molecular polarity [11, 12, 14, 19-21].

To illustrate the wetting behavior of ILs on different surfaces, an ample number of experimental and theoretical simulation studies have been conducted[3, 4, 7, 9, 10, 13, 16, 32-38]. Qualitatively the wetting behavior of ILs significantly depends on the relative interaction between dispersion and electrostatic forces, as reported by Rane et al. and Restolho et al. [16, 38]. Pereira et al. [15]elucidate the effect of hydrogen-bonding ability of ILs to wet polar and non-polar surfaces. The effect of alkyl chain and surface tension on the wetting behavior of ILs on the graphite surface was investigated by Bordes et al. [39] and Bhattacharjee et al.[5] wherein the contact angle of the droplet decrease with an increase in the length of the alkyl chain. The effect on the nanodroplet size on the contact angle as a function of surface interaction energy was stated by various authors[9, 10, 13, 32, 35, 37]. Guan et al.[9] reported viscosity of ILs play a decisive role in the nanowetting of IL droplets, and the wettability of ILs nanodroplet decreases with the increase of the viscosity. Malali and Foroutan[13] reported that [BMIM][PF$_6$] on the crystallographic titanium dioxide surface possesses a cation rich layer due to the strong interaction between the ionic liquid and the surface. A significant role of charged moieties of the cation is observed when the adsorption and wetting of ILs on the boron nitride surface are concerned, as stated by Ghalami et al.[35]. Although the above mention parameters dominated the wetting and interfacial behavior of pure ILs, the inherent characteristics in the wetting behavior of ILs with a small amount of water are still not been explored yet to the best of our knowledge.

Recent studies on the microenvironment of water and IL, both experimentally[40-44] and computationally[21, 45-53], illustrate the presence of a small amount of water, leading to form isolated water molecule in which significant interaction between anion and water dominates. However, with an increase in water concentration, water molecules start to interact with each other to form clusters and lead to a strong network of water molecules at higher water concentrations. Zhong et al.[21] reported 1-butyl-3-methylimidazolium tetrafluoroborate [BMIM[BF$_4$] at various mole fraction of water and found, at the mole fraction of water $x_w$ <0.2, most of the water molecules are isolated in the polar cation−anion network in ILs; with an increase in the mole fraction $x_w$ from 0.2 to 0.8, large water clusters are formed, and at $x_w$ > 0.8, ionic liquids show a deliberate degree of aggregation and the later, the system transforms from IL-rich to water-rich mixture. Interestingly, the threshold value of water, which is ($x_w = 0.8$), was also found for 1-ethyl-3-methylimidazolium ethylsulfate [C$_2$MIM][EtSO4]/water mixture in a recent molecular dynamics simulation study by Bernardes et al.[54]. However, others[21, 55] reported significant water clustering at $x_w > 0.4$ for [BMIM][BF$_4$]/water. On the other hand, Niazi et al.[56] predicted insignificant water clustering in chloride and acetate-based IL/water such as butyl-3-methylimidazolium chloride [BMIM][Cl], 1-ethyl-3-methylimidazolium acetate [EMIM][Ac] and 1,3-dimethylimidazolium dimethyl phosphate [DMIM][DMP] mixtures relatively at moderate water fraction, followed by an incremental rise in cluster growth at $x_w > 0.7$ and the transition from IL rich phase to an aqueous solution in all three ILs is noticed at water concentrations at $x_w \sim 0.7$.

Thus, from the above mention studies, molecular dynamics has prevailed as a powerful tool for an in-depth understanding of the water-ILs system at nanoscale [47, 56-59]. ILs behavior in aqueous solutions has been subject to numerous investigations [58-61]. Depending on the alkyl chain length, concentration and selection of anion/cation, self-organization, aggregation, and phase separation behaviors, clusters, are some of the traits that have also been reported when water is present in the ILs [44, 62 30, 63-67]. Similar changes at the solid-liquid interface may cause differences in macroscopic wetting, affecting both the contact angle and spreading kinetics.

Wetting characteristics of ionic liquid in aqueous solution subject to a fewer number of studies [68-70]. The small amount of water with ionic liquid has been reported to have a negligible effect on the electrowetting behavior of ionic liquid [BMIM][BF$_4$] as shown by Paneru et al.[69] However, significant impact of water concentration on the interfacial and wetting behavior is reported by Wang et al.[71] wherein interaction of 1-alkyl-3-methylimidazolium bis(trifluoromethanesulfonyl)-imide, [RMIM][NTF$_2$] ions with water molecules through hydrogen bonding is suspected of disrupting the IL molecular layering and precursor film formation. The formation of a precursor film of the thickness ($10^{-6} \sim 10^{-10}$ m depending on the system) is usually found when macroscopic drop spreads on a solid surface [72-74]. The effect of water on macroscopic spreading behavior is expected to inform the design of IL-based fluidic technologies and assist in understanding complex static and dynamic wetting behavior of analogous complex fluids.[75] Here, for the first time to the best of our knowledge, MD simulations are performed to reveal the wetting behavior and intrinsic mechanism of ILs to disclose the effect of water concentration on the interfacial behavior of hydrophilic imidazolium-based IL [EMIM][BF4] on the graphite surface.

## 2. Computational Model and Methods

MD simulation performed on LAMMPS[76] to study the wetting and the interfacial behavior of hydrophilic IL [EMIM][BF4]. Non-bonded interaction parameters of the ILs were described by the all-atom optimized potentials for liquid simulations (OPLS-AA) force field [77-79], which has been used to acquire the structures and liquid properties of ILs efficiently [80-82]. The rigidly extended simple-point-charge model (SPC/E) water was used to describe the water molecules[83]. 12−6 LJ potential parameters for the carbon−carbon interactions of graphite surface was taken from Werder et al.[84]. Mixing rules were applied to cross LJ interactions. ILs pairs initially distributed randomly in a cubic box using Packmol software[85]. The cubic box of ILs is then placed just above the graphite surface (within the distance of 10 Å), as shown in **Figure 1**. During the equilibration period, the ILs gradually turned into a droplet shape on the surface. The dimension of the graphite surface and the water concentration varied from 0 to 70 wt.% for which the equivalent numbers of ion pairs (IPs) are listed in **Table 1.** The positions of graphite sheets were fixed during the simulations. The cutoff distance for intermolecular interaction used is 12 Å, and the (PPPM) method with an accuracy of 0.0001 applied for long-range interactions[86]. For both IL and water, the SHAKE [87]algorithm is used to fix the bond length and angle. A Nose−́ Hoover [88] thermostat is used to maintain the system temperature with a relaxation constant of 1.0 ps. The equation of motion was integrated using the Verlet-velocity algorithm with a time step of 1 fs. The height of the simulation box is 200 Å to avert the interaction among periodic images of the drop. Before placing on the graphite, the box of IL and water was equilibrated in NVT simulation at 300 K for 5ns, which later turned into a spherical droplet. Another 20-30 ns was simulated for further equilibration of the droplet on the surface with a time step of 1 fs. Details of the contact angle measurement can be found in our previous publications[5].

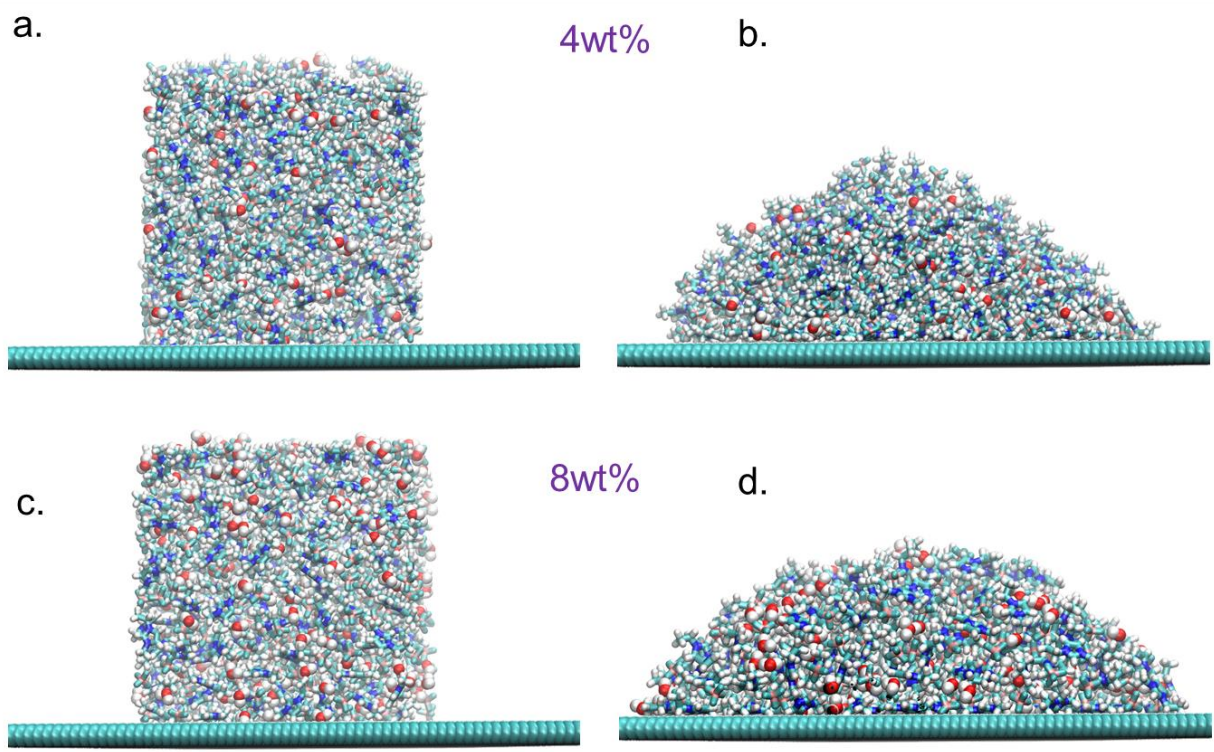

**Figure 1.** (a) Snapshot of Initial and (b) Final equilibrate shape of 4 wt.% water added in 250 IP of [EMIM][BF$_4$] and (c) Snapshot of Initial and (d) Final equilibrate shape of 8 wt.% water added in 250 IP of [EMIM][BF$_4$] on the graphite surface.

**Table 1.** The weight percentage of water added for fixed 250 ion-pair of IL [EMIM][BF$_4$]

| Weight % of water | Water molecules | $x_w$ (mole fraction of water) | Sheet-dimension $(L \times W)$ nm$^2$ |
|---|---|---|---|
| 2 | 55 | 0.18 | 20.418 × 20.448 |
| 4 | 110 | 0.30 | 20.418 × 20.448 |
| 8 | 220 | 0.46 | 20.418 × 20.448 |
| 20 | 550 | 0.68 | 20.418 × 20.448 |
| 30 | 825 | 0.77 | 20.418 × 20.448 |
| 50 | 1375 | 0.85 | 20.418 × 20.448 |
| 70 | 1925 | 0.89 | 20.418 × 20.448 |

3. **Results and Discussion**

**3.1 Equilibrium of droplet:**

The equilibration of the droplet on the graphite surface is traced through the graphite-droplet interaction energy (E$_{inter}$) to be more or less constant over time. The evolution of solid-fluid interaction energy of the droplet for the two systems, *i.e.,* 2 wt.% and 70 wt.% of water are shown in **Figure 2.** Here, we have normalized the interaction energy of the droplet with respect to the total number of molecules. To attain the equilibrium state for 2 wt.% is taken significantly more time (≈ 27 ns) as compared to 70 wt.% (≈ 11 ns) **(see Figure 1).** This is due to the fact that the 2 wt. % water-IL system has a slow dynamic in nature (almost act as the pure IL, high viscous in nature). At low water content, the dynamics of the system mainly depends on the viscosity of the neat IL, whereas at higher water content, water cluster dominate the system and accelerate the system dynamics.[89]

Primarily the cation of the droplet takes a longer time to attained equilibrium due to large size, reordering at the vicinity of the surface, and high interaction with a graphite surface. Conversely, at the higher percentage of water, the system has been diluted, and normalized interaction energy decreased significantly, as shown in **Figure 2b**. Once the system reached equilibrium, another 10 ns is used for a production period of the system to compute the contact angle, density profile, and orientation profile of ILs droplet on the graphite surface.

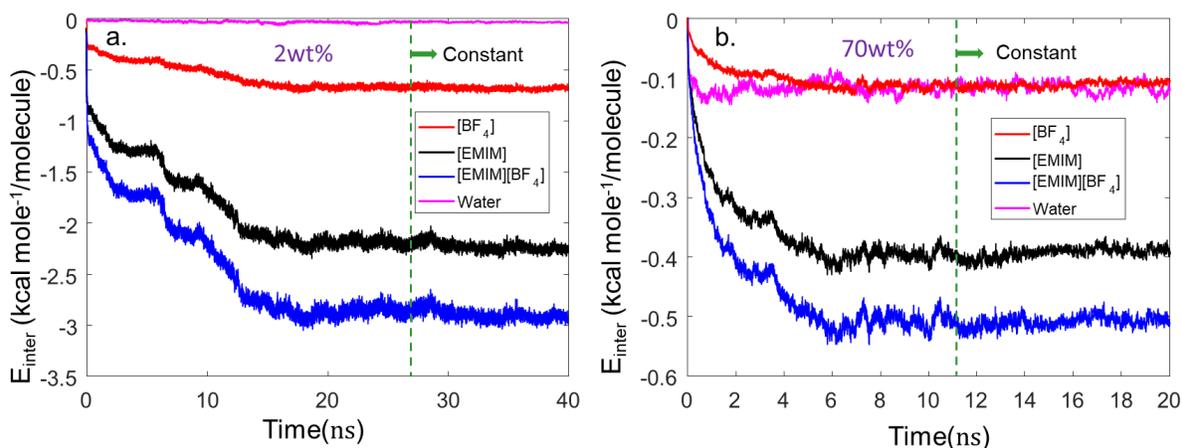

**Figure 2.** Evolution of graphite-IL interaction energy for (a) 2 wt. % [EMIM] [BF$_4$] (b) 70 wt.% [EMIM][BF$_4$]

**3.2 Contact angle**

To understand the effect of water concentration on the wettability of the IL droplet on the smooth graphite surface, we calculate the contact angle of the droplet at a different weight percentage of water, as shown in **Figure 3**. We use the atomic resolution of the droplet to determine the contact angle of the IL droplet on the graphite sheet of the droplet. Initially, the contact angle of the droplet slightly decreases while adding water up to 20 wt.%, then onwards (i.e., beyond 20 wt.% ), the contact angle starts to increase, as shown in **Figure 3**. Please note, here the value of the contact angle for pure IP ([EMIM][BF$_4$]) and aqueous IL (i.e., beyond 70 wt.%) are taken from our previous publications [10] and [90]

Experimentally, The decrement of contact angle was also reported by Wang [91] on the mica surface, where increasing the water concentration of IL [BMIM][BF$_4$] causes a decrease in contact angle value. It was presumed that only water concentrations higher than ($x_w$ = 0.21) could force this IL to rearrange its internal order, which subsequently changes the contact angle. As water concentration is increased further, more ions of [BMIM][BF4] are solvated by water, and the contribution of water to its macroscopic wetting behavior is more significant. Finally, at high concentrations of water, the [BMIM][BF4] will behave as water with a small amount of IL contamination. [92, 93].

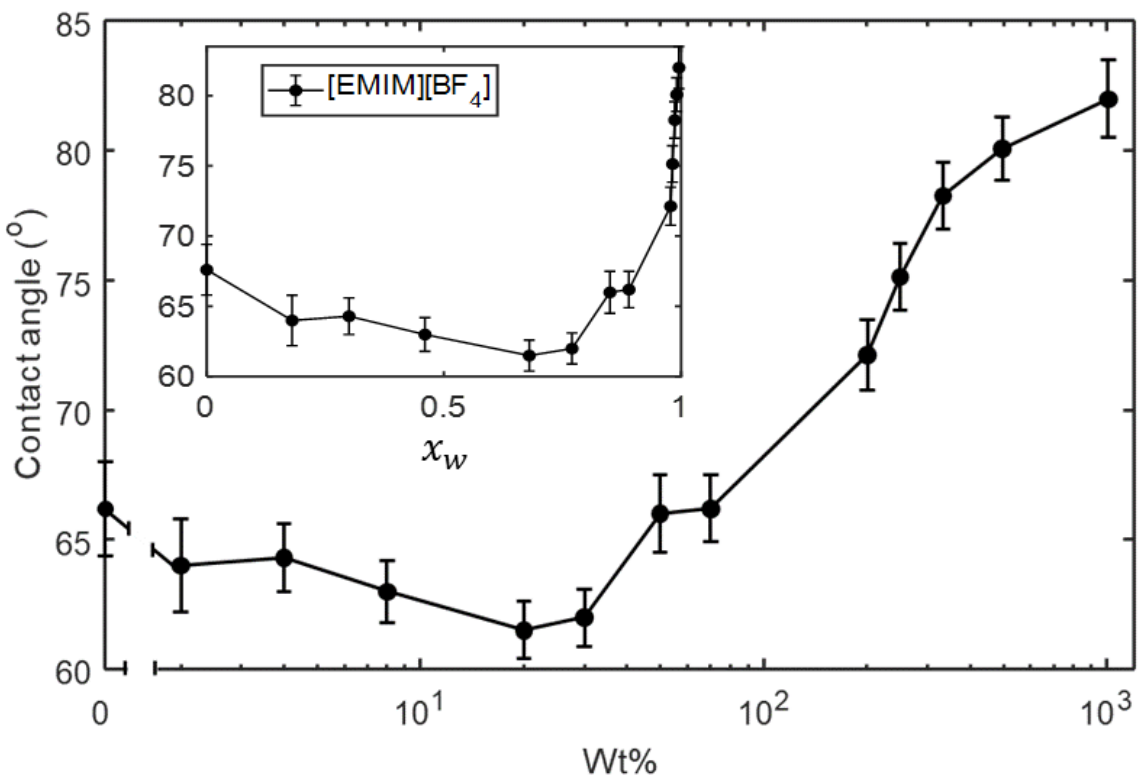

**Figure 3**. Variation of contact angle of the droplet with wt. % of the water in [EMIM][BF$_4$] on the logarithmic scale. The inset image shows the contact angle w.r.t. water mole fraction ($x_w$). Here all the contact angle value reported are calculated as averages over the last 5-6 ns during the production run after the droplet attains equilibration. Error bars represent the standard deviation (SD) of the contact angle.

### 3.3 Density Profile

Further, the wetting behavior of IL with water on a smooth graphite surface is characterized by a density contour profile. We used a block sum average method to calculate the density profile of the equilibrium droplet (after equilibration period) for 5 ns at 0.01 ps intervals. Density distribution of IL and water within the droplet for different wt. % of water are separately shown in **Figure 4a** and **Figure 4b,** respectively. It can be observed from the IL density **(Figure 4a)** profile that pure IL droplet has multiple peaks (multi-layer of IL), which are dissolved (vanished) as the concentration of water increases, except the first peak (i.e. near the surface). In particular, at low concentrations of water ($< 4$ wt.%), the water molecules are distributed uniformly across the droplet without any noticeable association of water molecules, i.e., water molecules are preferentially in isolated form. At the 8 wt.% of water, the water molecules start to accumulate above the $1^{st}$ layering of the IL. Hence, the aggregation of water molecules is observed (i.e., $Z \approx 5\text{Å}$), which is associated with the $2^{nd}$ layer of the droplet from the surface (see Figure 4b-iv, cyan stripe at $Z \approx 5\text{Å}$). Interestingly, we do not observe much water molecules in the $1^{st}$ layer (i.e., $Z < 3.35\text{Å}$), which is due to the strong ordering of cations and anions of IL molecules near the surface to maximize the surface interaction with the IL molecules.

Additionally, an increase in the water concentrations (20 wt.% and 30 wt.%), the degree of association of water molecules increases, and start forming of the water network and disrupt the IL layering as can be seen in **Figure 4(a) and 4(b).** Finally, a continuous phase of IL-water solutions appears at the bulk of the droplet beyond 50 wt. % of water. However, at these concentrations, we still observe intense layering of IL molecules near the surface, whereas intense layering of water molecules is found in the $2^{nd}$ layer (i.e., $Z \approx 5\text{Å}$) from the surface.

For further illustration, the density profile of the droplet along the z-direction is analyzed using a cylindrical section from the droplet of radius 20 Å perpendicular to the graphite sheet, as shown in **Figure 5.** Here, it can be noticed that the multi-peak curve (wavy in nature) gets smooth/filter while adding water on it (except the first layer). Due to intense interaction energy between the sheet and IL molecules, the first layer does not get affected. Conversely, water density uniformly distributed (no multiple layers observe) within the bulk of the droplet, and a dense layer of water molecules is appeared next to the IL layer near the surface at higher concentrations, as shown in **Figure 5.** With the addition of water, the interaction energy between cation and anion has been enfeebled, and hence the contact angle decreases (upto 20 wt. %). However, at a higher weight percentage of water (above 30 wt.%), water molecules dominate the system due to the strong association of water molecules, which can be examined through cluster formation within the droplet, discussed in the next section and hence, the contact angle of the droplet increases with further increase in water concentration. It can also be noticed that the vapor-liquid interface of water molecules crosses the vapor-liquid interface of IL molecules at higher concentrations (in between 30 wt.% to 50 wt. %) that also indicates the transition between IL rich to the water-rich system. We have also investigated the solid-liquid interface through separate simulation **(shown in supporting information Figure S1**) where the IL-water mixture is placed on the surface in a rectangular box. The z-density profile shows the same behavior as in the case of the spherical droplet **(shown in supporting information Figure S2).**

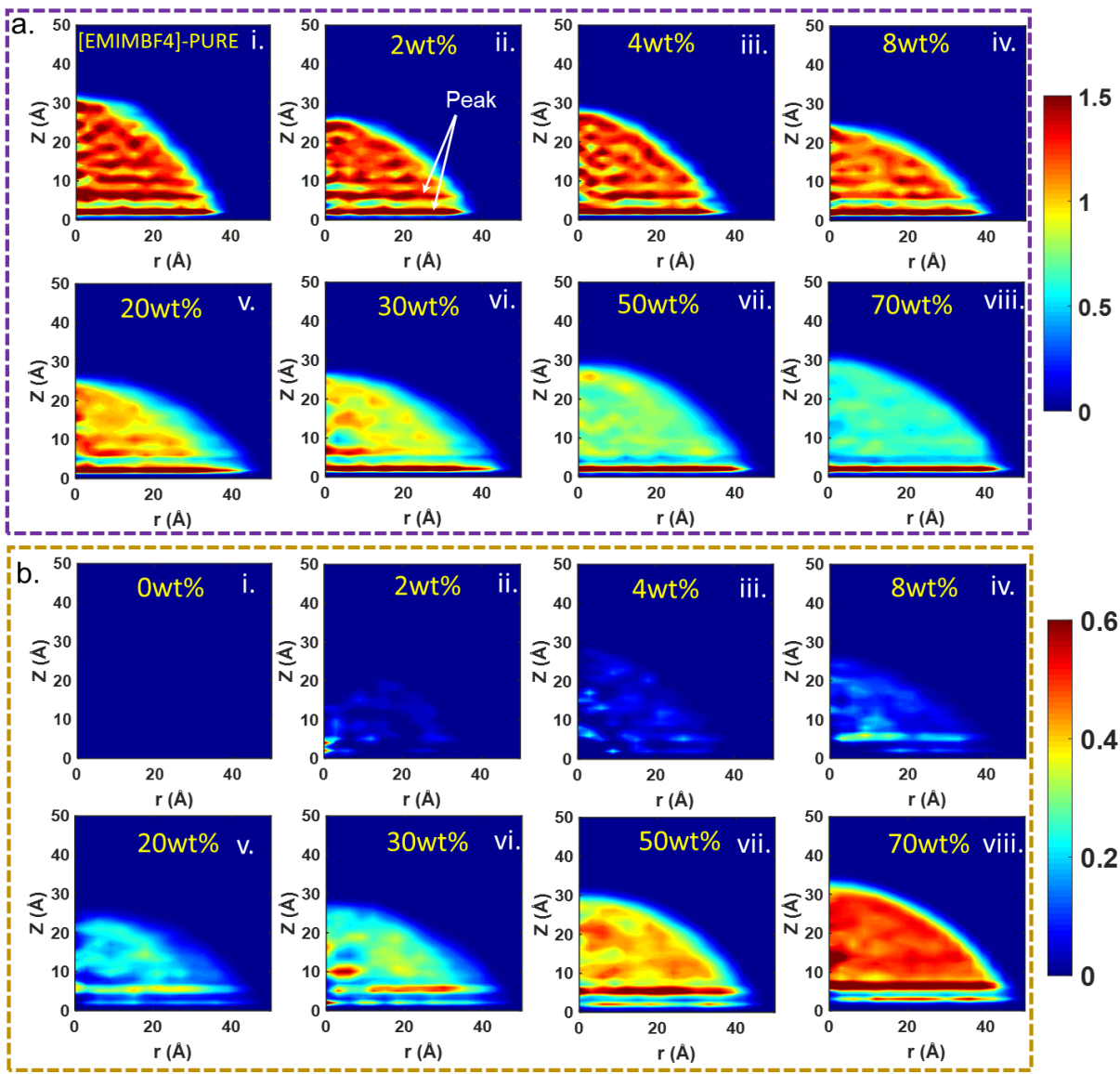

**Figure 4.** Density distribution of (a.) IL and (b.) water concentration at different weight percentages added in IL.

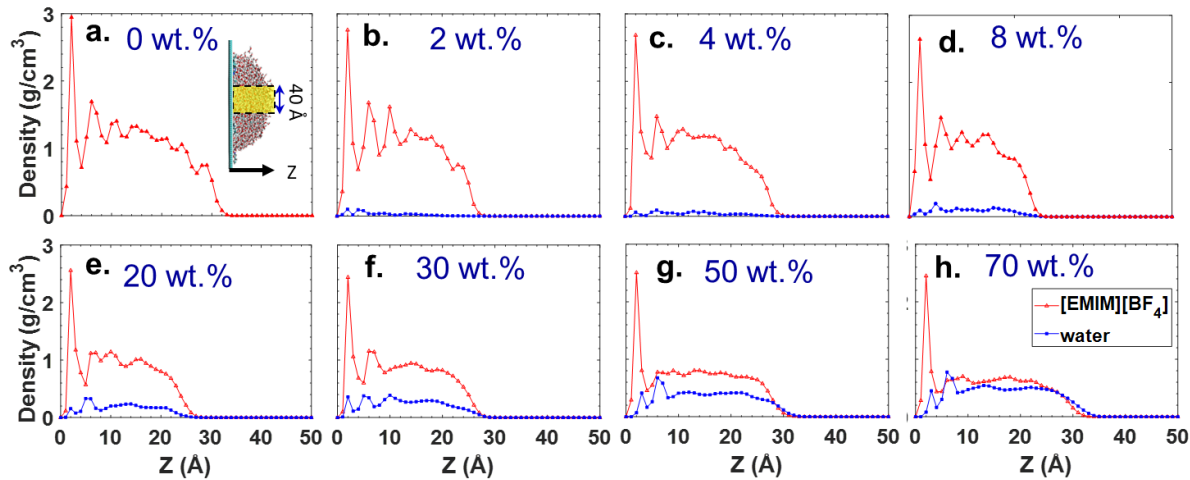

**Figure 5.** Z-direction density profile for IL and water for different weight percentages of water.

## 4. Water clustering within droplet:

The density contour across the droplet reveals that at low water concentration, the water molecules distributed uniformly within the droplet without affecting much on the IL layering. With increasing the water molecules, the layering of IL molecules gets smooth except the first layer near the surface. To understand the transition between IL rich to the water-rich system, we examine the cluster formation within the droplet with the addition of water molecules. **Figure 6** shows the cluster size distribution normalized with the total number of water molecules present in the system at different concentrations of water. At low concentration (up to 8 wt. %), most of the water molecules are found either in isolated form or in a small cluster (in the range of 1-12 molecules) within the droplet **(see Figure 7) w**ith an increase in water concentration, the size of the cluster increases. For example, at 30 wt.% of water, the cluster size is uniformly distributed across the number of water molecules present in the system, shown as a green dotted line in **Figure 6**. This indicates the formation of a water network within the droplet. Further, an increase in water concentration (see 50 wt.% in Figure 6), water molecules suddenly move to larger cluster sizes and form a strong water network within the droplet and disrupt the IL layering. Therefore, the transition between IL rich to the water-rich system can be found in between 30 to 50 wt% of water, and hence, the contact angle of the droplet is found to increase after the transition.

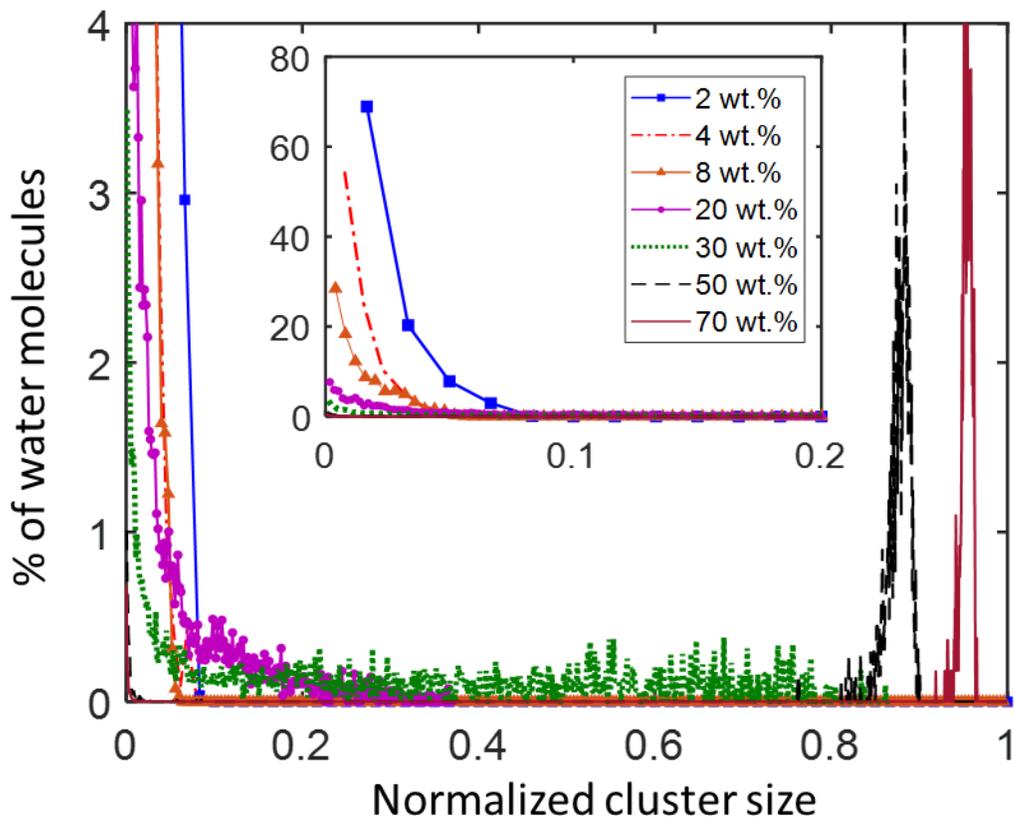

**Figure 6.** Cluster size distribution normalized with the total number of water molecules present in the system at different concentrations of water

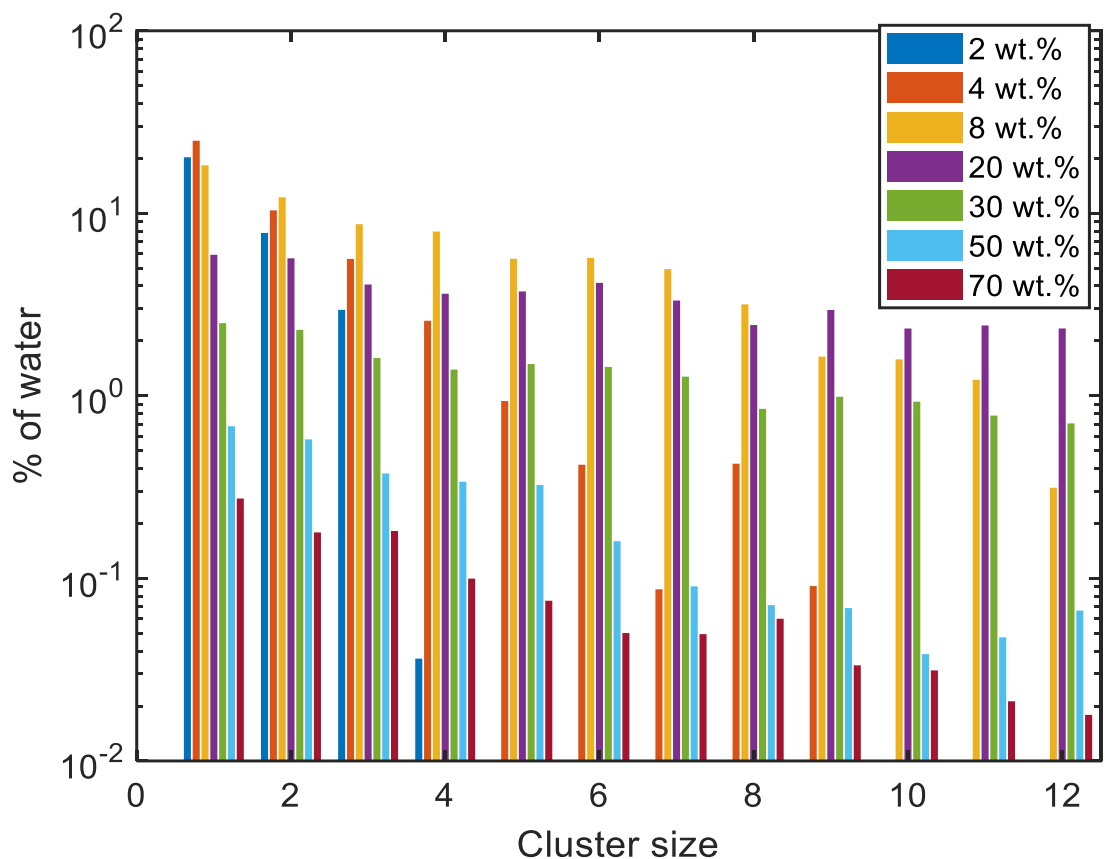

**Figure 7.** Size distributions of water cluster at various water mole fractions.

## 5. Hydrogen Bond Analysis

It was illustrated by many studies[52, 94-96] that the interaction between anion and water plays a significant role when a small amount of water is absorbed in IL. Both experimental[97] and computational investigations[21, 50] elucidate specific water effects on the microstructure of the imidazolium-based IL, where the anion [BF$_4$] is strongly hydrogen-bonded to water molecules at low water concentrations. For example, Toshiyuki et al.[52] investigated [EMIM][BF$_4$] and water within a range of mole fraction $0.09 \leq x_w \leq 0.34$ through ATR-IR spectroscopy and revealed that water molecules are mainly hydrogen-bonded individually to the anions at $x_w \leq \sim 0.2$; while the molar fraction $x_w > \sim 0.3$, water molecules are hydrogen-bonded among themselves in the solutions. This is due to the formation of aggregates by water molecules through hydrogen bonding, which further weakens the interactions between cation [EMIM] and water as well as anion [BF4] and water. Similarly, Jing Zhou [98] investigated 1-butyl-3-methylimidazolium acetate ([BMIM][Ac]), 1-butyl-3-methylimidazolium tetrafluoroborate [BMIM][BF$_4$] and 1-butyl-3-methylimidazolium bis(trifluoromethylsulfonyl)imide [BMIM][TF$_2$N] with entire range ($xw = 0$, 0.1, 0.3, 0.5, 1.0) of water and found weakening of HB interaction between cation and anion with the increase of water content. Because the water molecules form HB with anions by destroying the HB network between cations and anions. On the other hand, Porter et al.,[99] perform MD simulation of hydrophilic IL [EMIM][BF$_4$] as well as hydrophobic IL [EMIM][NTF$_2$] at low water concentration ($xw < 0.1$) and reveals that the water molecules with varying cluster sizes are typically each hydrogen-bonded to two anions, which is in agreement with the conclusions of the experimental study of Cammarata et al. .[94]. Similarly, Koishi et al. [47] have also observed strong interaction between water-anion in the case of [BMIM][BF4] at low water concentrations.

Correspondingly, we have also estimated HB between anion-water and water-water to understand the nature of droplets with water concentration. We measured HB between anion-water and water-water based on geometric criteria [47], also shown in (**Figure 8**).

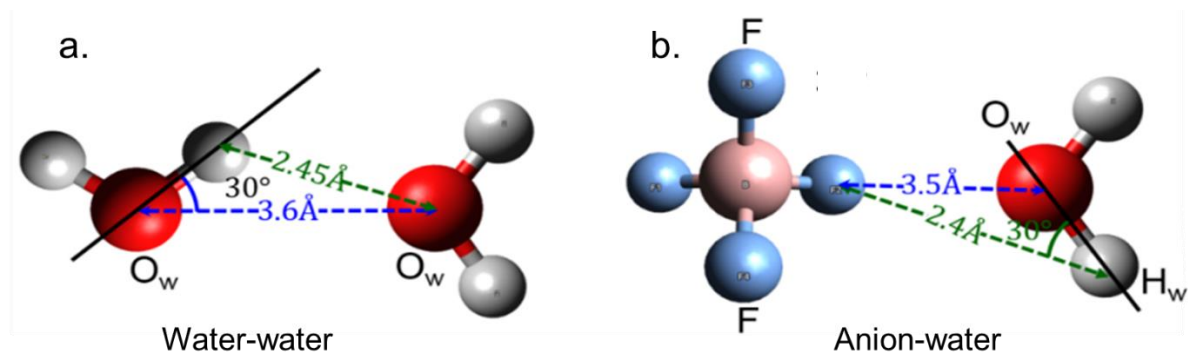

**Figure 8.** Hydrogen bond (HB) geometric criteria (a) for water-water (b) for anion-water.

**Figure 9 and 10** shows the distribution of the hydrogen bond between water-water and water-anion within the droplet. We have also reported the total average number of HBs for the whole system normalized by the total number of water molecules. We can observe that the number of HBs between water-water increases, and water-anion decreases with the increase in water concentration. Furthermore, water-water and anion-water HBs are both uniformly distributed in the droplet due to the hydrophilic nature of IL molecules. More importantly, at 30wt%, water-water and anion-water both have the same ⟨HB⟩value (i.e.$\langle HB.\rangle_{w-w@\,2wt\%} = \langle HB.\rangle_{a-w@\,2wt\%} = 0.83$), and beyond 30wt%, water-water HB dominates over anion-water HB. Hence, the contact angle of the droplet increases beyond 30 wt. % as water-water interaction drives the system. In addition, we have also examined the hydrogen bond profile normal (i.e., along $Z$ direction) to the surface to analyze the HBs near the surface and the bulk of droplet, as shown in **Figure 10**. We can observe that near the surface HB between water-water is less compared to the bulk of the droplet ($\langle HB.\rangle_{@sheet} < \langle HB.\rangle_{@bulk}$, See **Figure 10**), whereas the HB between anion-water is more near the surface compared to the bulk of the droplet. This observation suggests that the anion-water hydrogen bonding is more preferable than the water-water hydrogen bonding near the surface.

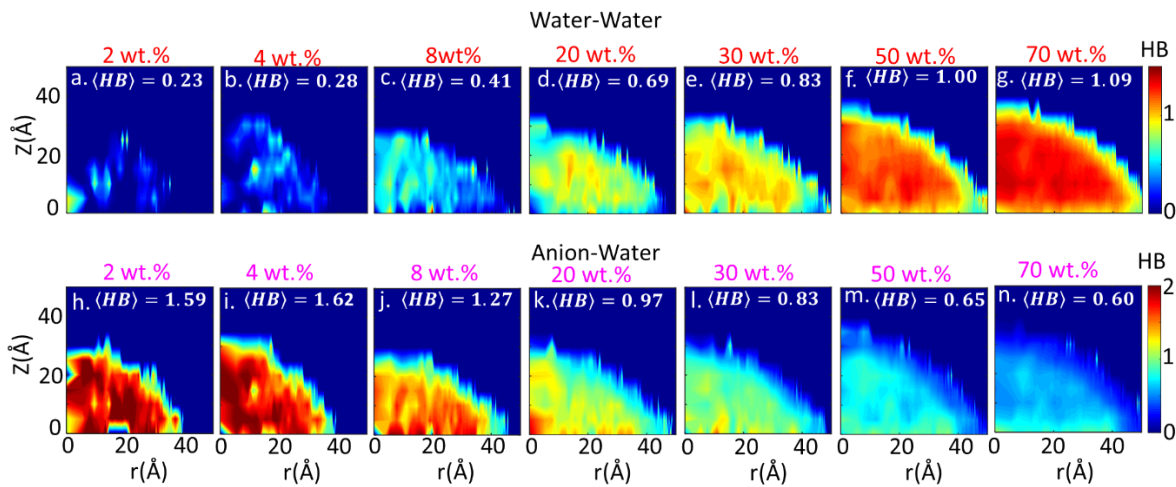

**Figure 9.** Hydrogen bond distribution in the droplet (a-g) water-water hydrogen bond. (h-n) Anion-water hydrogen bond for different weight percentages of water.

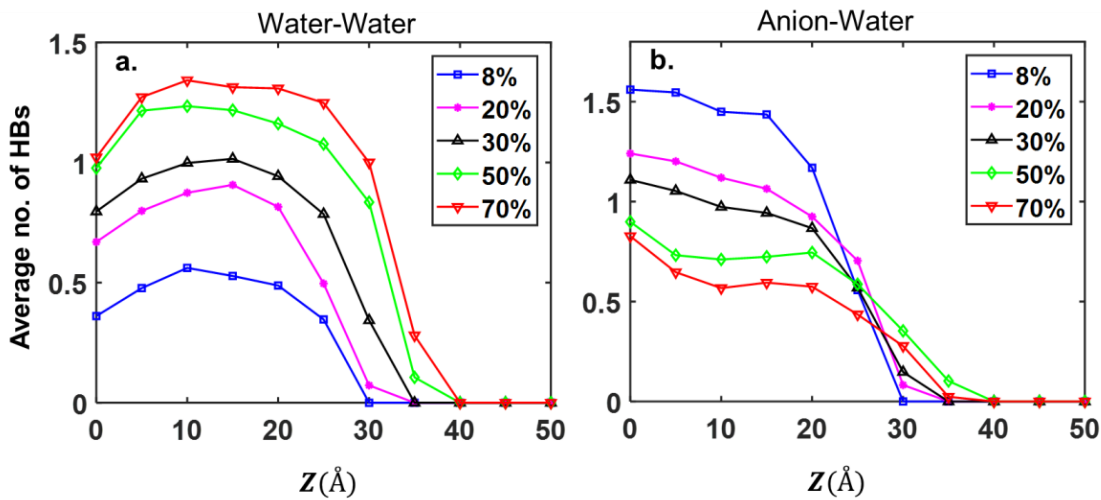

**Figure 10.** The plot of the average number of hydrogen bonds w.r.t. z-direction (a.) Water-water HB. (b.) Anion-water HB.

**Conclusions**

The presence of water as an impurity in ILs can play a different role than in the aqueous phase of IL. In summary, we conducted MD simulations to study the effect of water on the wetting behavior and the interfacial structure of [EMIM][BF4] IL droplets on graphite surfaces over the wide range of water percentage (0 wt.%-70 wt.%). We quantified several important parameters, such as the interaction energies between the graphite and IL molecules, contact angle, density distributions, cluster formation, and the number of formed HBs for the different weight percentages of water molecules. It has been found initially contact angle of the droplet slightly decreases while adding water up to 20 wt.%, and then onwards, the contact angle starts to increase (water molecules dominated). The density profile normal to the surface shows that there is not much effect on the layering of IL molecules at low concentration of water, whereas, at a high concentration of water molecule, the layering of IL molecules gets disrupted except the 1$^{st}$ layer near the surface. The degree of association of water molecules increases with an increase in water concentration, as observed from cluster analysis. We found that the IL rich regime of the droplet transforms into a water-rich regime above 30 wt.% of water. Hydrogen bond analysis reveals that the water-water hydrogen bonding dominates over the anion-water hydrogen bond above 30 wt.% water, which is responsible for the increase of contact angle of the droplet above this concentration. We believe that our study can provide an in-depth understanding of the wetting behavior of IL in the presence of a small amount of water, which could be useful in many industrial applications.

# Supporting Information

## Quantification of the Impact of Water on the Wetting Behavior of Hydrophilic Ionic liquid: A Molecular Dynamics Study

*Sanchari Bhattacharjee, Sandip Khan* *

Department of Chemical & Biochemical Engineering

Indian Institute of Technology Patna

Patna, India- 801103.

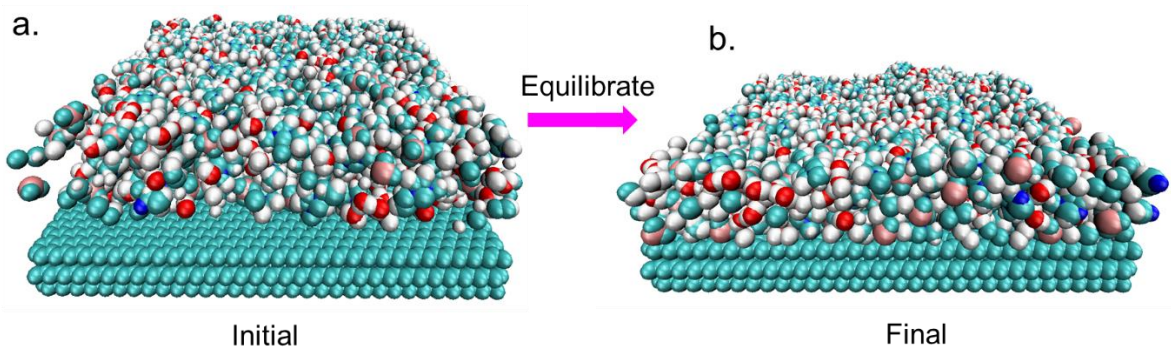

**Figure S1.** (a) Snapshot of Initial and (b) Final equilibrate shape of 20wt.% water added in 250 IP of [EMIM][BF$_4$].

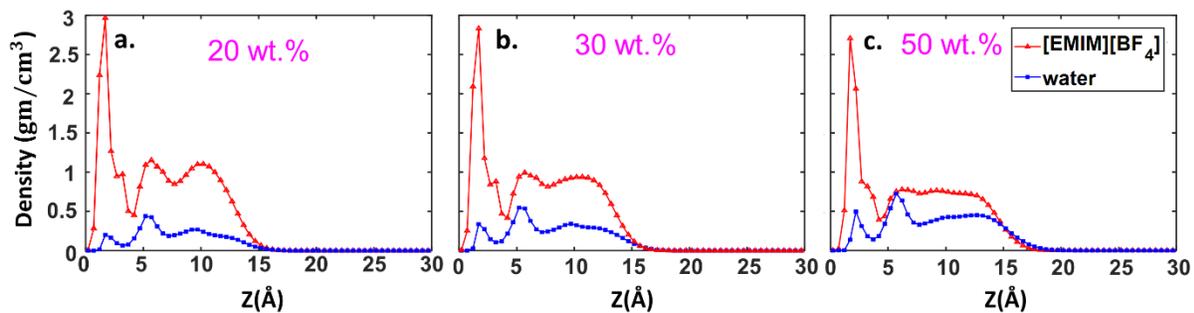

**Figure S2.** Z-direction density profile for IL and water for different weight percentages of water for the IL-water mixture system placed on the surface in a rectangular box.

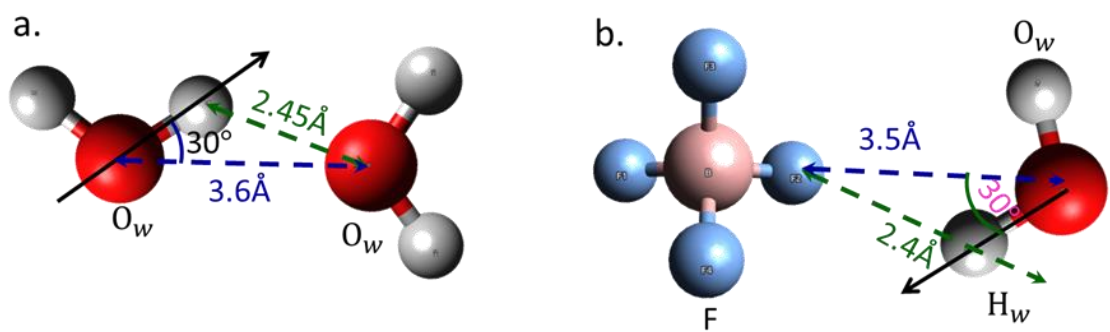

**Figure S3.** Hydrogen bond (HB) geometric criteria (a) for water-water (b) for anion-water.

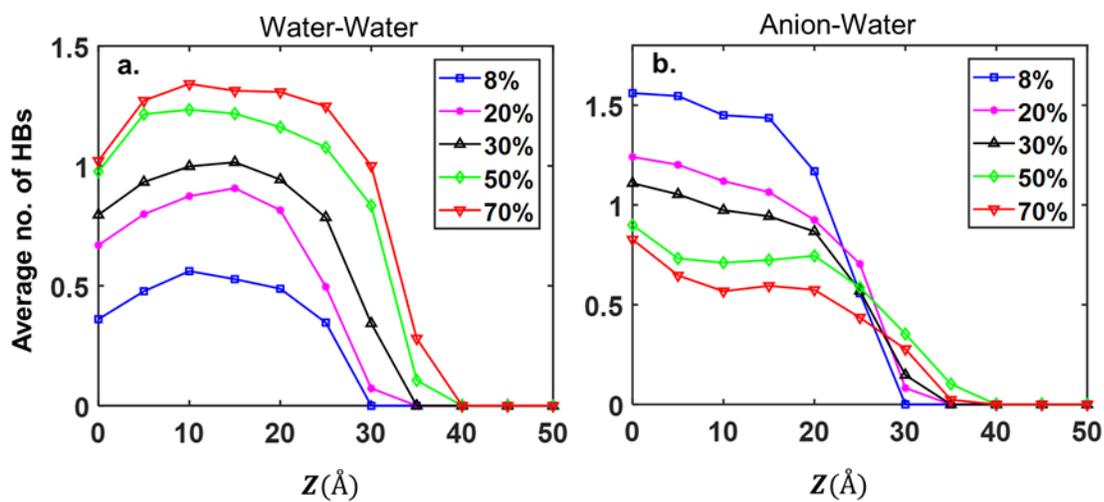

**Figure S4.** The plot of the average number of hydrogen bonds w.r.t. z-direction (a.) Water-water HB. (b.) Anion-water HB.